\documentclass[a4paper,11pt]{article}
\usepackage{pos}

\usepackage[]{caption}
\usepackage{subcaption}
\usepackage{amsmath,amssymb}

\title{Search for Astrophysical Neutrino Transients with IceCube DeepCore}
 \ShortTitle{Neutrino Transients with IC-DeepCore}

\author{The IceCube Collaboration \\{\normalsize \normalfont(a complete list of authors can be found at the end of the proceedings)}}

\emailAdd{cjchen@mail.icecube.wisc.edu}
\emailAdd{pdave@gatech.edu}
\emailAdd{itaboada@gatech.edu}

\abstract{DeepCore, as a densely instrumented sub-detector of IceCube, extends IceCube's energy reach down to about 10 GeV, enabling the search for astrophysical transient sources, e.g., choked gamma-ray bursts. While many other past and on-going studies focus on triggered time-dependent analyses, we aim to utilize a newly developed event selection and dataset for an untriggered all-sky time-dependent search for transients. In this work, all-flavor neutrinos are used, where neutrino types are determined based on the topology of the events. We extend the previous DeepCore transient half-sky search to an all-sky search and focus only on short timescale sources (with a duration of $10^2 \sim 10^5$ seconds). All-sky sensitivities to transients in an energy range from 10 GeV to 300 GeV will be presented in this poster. We show that DeepCore can be reliably used for all-sky searches for short-lived astrophysical sources.

\vspace{4mm}
{\bfseries Corresponding authors:}
Chujie Chen$^{1}$, Pranav Dave$^{1*}$, Ignacio Taboada$^{1}$\\
{$^{1}$ \itshape Georgia Institute of Technology}\\
$^*$ Presenter

\FullConference{37$^{\rm{th}}$ International Cosmic Ray Conference (ICRC 2021)\\
		July 12th -- 23rd, 2021\\
		Online -- Berlin, Germany}

}

\begin{document}
\maketitle

\section{Introduction}\label{sec:info}

Astrophysical neutrinos, as one important messenger in multi-messenger astronomy, are thought to come from several types of celestial objects and phenomena. Several analyses have been performed within IceCube to search for origins of high-energy neutrinos. In this proceeding, we mainly focus on low-energy neutrinos and their possible origins, i.e. transient astrophysical point sources. 

Gamma-ray bursts (GRBs) are mostly extragalactic phenomena that produce extraordinarily bright emission of gamma rays lasting between 0.1 and 1000 seconds. The fireball model \cite{rees1992relativistic} is one the most widely accepted model describing gamma-ray bursts, although GRBs are not completely understood yet. In this model, a compact rotating object, either a black hole or a short lived neutron star, powers the emission of jets as it undergoes rapid accretion. The jets, that are accelerated to relativistic speeds, are oriented along the object's axis of rotation. Materials in the jets will form sub-shells and result in internal shocks where particles are accelerated. The accelerated protons are responsible for the production of a neutrino flux while electrons produce gamma-rays through synchrotron radiation. For choked gamma-ray bursts (choked-GRBs), material in jets cannot breach the surrounding stellar envelope because of insufficient energy or dense surrounding envelope, while neutrinos can leave the object and have a chance to be observed by IceCube. The model proposed by Razzaque, M{\'e}sz{\'a}ros and Waxman \cite{razzaque2004tev}
and further developed by Ando and Beacom \cite{ando2005revealing} for choked-GRBs predicts that the energy spectrum has a relatively soft spectral index ($\gamma \sim 3$) due to dominating pion decays. The breaks in the energy spectrum at energies below IceCube’s optimal energy range makes the use of low-energy neutrinos to search for those sources possible. Another model that is similar to the fireball is the subphotospheric model. In this model, protons in the relativistic jets decouple from the neutrons and reach higher Lorentz boost factor than neutrons. Particles like pions are produced due to the inelastic collision between protons and neutrons resulting in a predicted energy range from 10 GeV to 100 GeV \cite{murase2013subphotospheric}.

\section{Datasets}\label{sec:info}

The dataset used in this analysis are IceCube events detected by the DeepCore sub-array. Those events have energies from a few GeV to $\sim$ O(10 TeV). Several cuts are applied in order to eliminate background atmospheric muons and noise-dominated events. At the final level, the data sample is dominated by atmospheric neutrinos. Downgoing atmospheric neutrinos are of the electron and muon flavors, but for upgoing events, the tau flavor also contibutes. In DeepCore, muon neutrinos create tracks, via the charged current interaction, while electron and tau neutrinos and muon neutrinos via the neutral current interaction produce cascades. The data sample used in this analysis contains both upgoing and downgoing events that are reconstructed using a hybrid track+cascade fit so it is suitable to do all-sky searches. More information about the dataset used in this work can be found in \cite{IceCube:2020qls} and \cite{icrc2021mlarson}.

\begin{figure}[h!]
\centering
\begin{subfigure}{.5\textwidth}
  \centering
  \includegraphics[width=\linewidth]{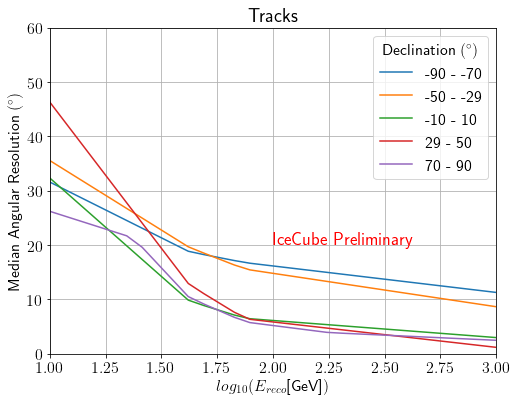}
  \caption{track angular error plot}
  \label{fig:sub1}
\end{subfigure}%
\begin{subfigure}{.5\textwidth}
  \centering
  \includegraphics[width=\linewidth]{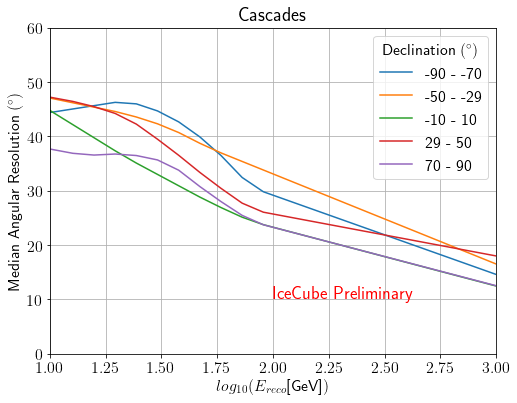}
  \caption{cascade angular error plot}
  \label{fig:sub2}
\end{subfigure}
\caption{Median spline per-event angular resolution for tracks (a) and cascades (b) as a function of logarithmic reconstructed neutrino energy at several declinations. Data are weighted according to the atmospheric spectrum.}
\label{fig:spline}
\end{figure}

The reconstructed track length of each event is used as particle identification (PID) to classify events as tracks or cascades. When the reconstructed track length is longer than 50m, the corresponding event is classified as track event, while cascade events have shorter lengths. The angular resolution plots shown in Figure \ref{fig:spline} using simulated data are weighted based on the expected energy spectrum for atmospheric neutrinos as described in \cite{honda2004new}. The results show that the angular resolution of tracks is better than that of cascades as in general they have longer lever arms of Cherenkov light within the detector. For both tracks or cascades, the higher the energy is, the smaller the median angular resolution is, as there are more digital optical modules activated and more information for the reconstruction to get more accurate resolution. For events at very low energies (10 GeV - 30 GeV), event topologies are poorly reconstructed. For track events having energies higher than 30 GeV, up-going events have better angular resolution than down-going events as optical modules are oriented downwards and more sensitive to photons coming upwards. Note that the behaviour above $\sim$ 300 GeV should be ignored due to the lack of statistics for both track and cascade events.

\section{Likelihood Model}\label{sec:info}

To identify astrophysical signal events among a vast background and thus localize the source, a time-dependent point source maximum likelihood method is, for the first time, used on these datasets. Information from events such as direction, arrival time and reconstructed energy is utilized in the likelihood model under certain hypotheses which allows us to obtain estimated parameters of the probability distribution to find potential clusters indicating sources. We use the unbinned maximum likelihood method with proper parameters for this untriggered time-dependent analysis. 

In this analysis, only spatial and temporal information are used. The signal probability distribution function (PDF) consists of a spatial and a temporal term for an event $i$.
\begin{equation}
    S_i(\vec{x_i}, t_i) = S_{space,i}(\vec{x_i}) \times S_{time,i}(t_i)
\end{equation}
Similarly, we have the background PDF,
\begin{equation}
    B_i(\vec{x_i}, t_i) = B_{space,i}(\vec{x_i}) \times B_{time,i}(t_i)
\end{equation}
For the signal PDF, $S_{space,i}$ is described using the Kent-Fisher distribution,
\begin{equation}
    S_{space,i} = \frac{\kappa_i}{4 \pi sinh(\kappa_i)} exp(-\kappa_i cos(|\vec{x}_{source} - \vec{x}_i|))
\end{equation}
where $\kappa_i = 1 / \delta_i^2$ in which $\delta_i$ is the angular uncertainty and $|\vec{x}_{source} - \vec{x}_i|$ is the angular separation between the direction of the event $\vec{x}_i$ and that of the hypothetical source $\vec{x}_{source}$. A Kent-Fisher distribution is used as, unlike a normal function, it is properly normalized over the surface of a sphere. The temporal PDF, $S_{time}$ is a uniform distribution to profile `box'-like clusters.

\begin{figure}[h!] 
\centering
\includegraphics[width=1.0\textwidth]{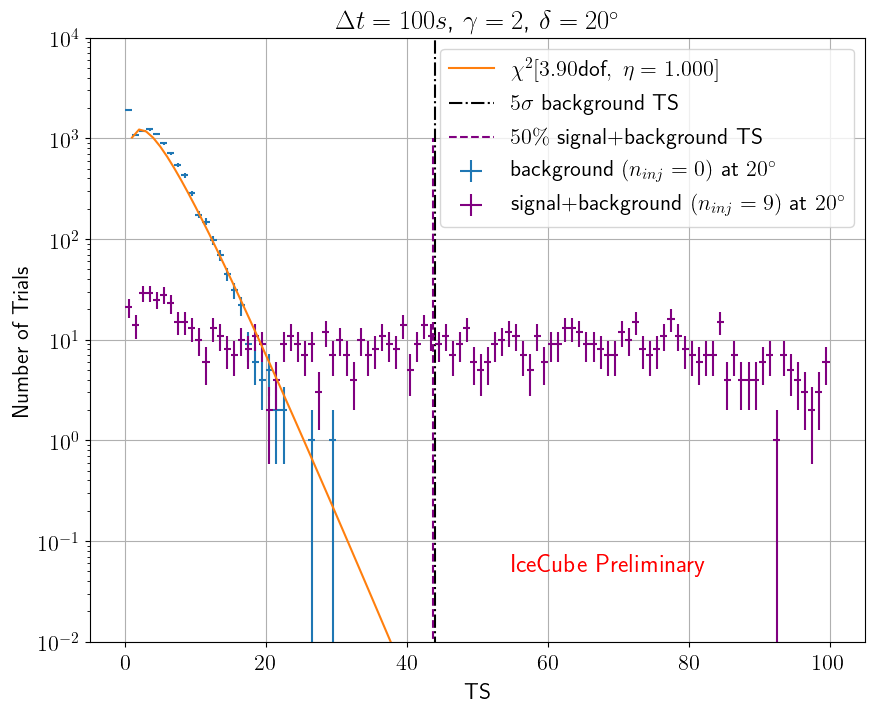}
\caption{Distributions of the maximized test statistic from 10k background trials (blue) and 1,000 signal injection trials (magenta) at $Dec = 20^\circ$. The dashed solid black line marks the location of a one-sided 5-$\sigma$ deviation while the dashed dark magenta line indicates the 50\% of the signal test statistic distribution. The injection is made assuming a $\gamma = 2$, Poisson mean $n_{inj} = 9$, and a flare width of $\sigma_w = 100 s$.}
\label{fig:TS_dist}
\end{figure}

For the background PDF, the spatial and temporal terms depend on the position of the event $i$. In this analysis, events are scrambled in right ascension, and thus a uniform distribution of events in right ascension can be assumed, but a more general representation looks like,

\begin{equation}
    B_i(\vec{x_i}, t) = B_{i,\text{declination}}(\vec{x_i}) \frac{B_{i,\text{right ascension}}(\vec{x_i})}{T} 
\end{equation}
where $T$ is the total livetime of the analysis. Thus, the probability of seeing an event $i$ given a time-dependent source hypothesis is
\begin{equation}
    P(n_s) = \frac{n_s S_i}{n_s + n_b} + \frac{n_b B_i}{n_s + n_b}
\end{equation}
where the $n_s$ and $n_b$ are the number of signal neutrinos and background neutrinos respectively. Thus, we have the likelihood function
\begin{equation}
    \mathcal{L}(\vec{x}_s, n_s, t_0, \sigma_w) = \prod_{i}  \left(\frac{n_s S_i}{n_s + n_b} + \frac{n_b B_i}{n_s + n_b}\right)
\end{equation}
to find the best fits of the number of signals $n_s$, the mean time of the source flare $t_0$, and the duration of the flare $\sigma_w$ at the hypothetical source location $\vec{x_s}$. We define the null hypothesis as the case for background only, $n_s = 0$, and the test statistic ($TS$) is defined as

\begin{equation}
\begin{aligned}\label{TS}
TS(\hat{\theta}) &= -2 \cdot  ln\frac{\mathcal{L}(n_s = 0)}{\mathcal{L}(\hat{\theta})} \\
&= 2 \cdot \sum_i ln\left [ \left(\frac{S(\vec{x}_i, t_i, E_i|\hat{\theta})}{B(\vec{x}_i, t_i, E_i|\hat{\theta})} - 1\right)\frac{n_s}{n_s + n_b} + 1 \right ]
\end{aligned}
\end{equation}
where the $\hat{\theta}$ stands for a set of parameters: the starting time $t_0$ and time duration of a signal cluster $\sigma_w$, and number of signal events $n_s$. Note that the per event angular uncertainties are weighted based on the selected energy spectrum, and thus, we have the energy term $E_i$ in the above equation.

\begin{figure}[h!] 
\centering
\includegraphics[width=1.0\textwidth]{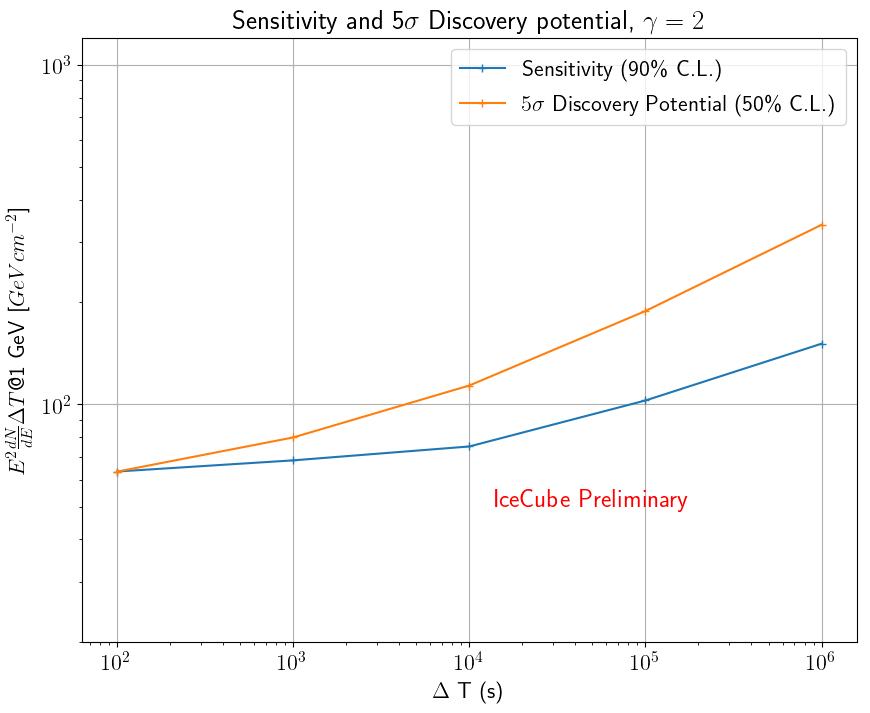}
\caption{Time-integrated sensitivity flux and $5\sigma$ discovery potential flux as a function of injected time width at $Dec = 0^\circ$. The $5\sigma$ background $TS$ is estimated using the Wilk's theorem. At longer timescales, the background increases, degrading the discovery potential.}
\label{fig:sens_dis_dt}
\end{figure}

\section{Results of the Likelihood Analysis}

Results from the untriggered time-dependent analysis are shown in this section. In this analysis, we maximize the $TS$ with respect to the fitting parameters $\hat{\theta}$. Experimental data scrambled by the time of arrival is used as background data to make events uniformly distributed within the livetime of the dataset. Since IceCube is located at the South Pole, scrambling the events in azimuth while keeping the declination distribution the same is adopted to get scrambled datasets. Injected signal events are generated corresponding to the assumed spectrum as the injection follows a Poisson distribution with the mean number of events $n_s$. The background $TS$ distribution at an example declination $20^\circ$ is shown in Figure \ref{fig:TS_dist}.

\begin{figure}[h!] 
\centering
\includegraphics[width=1.0\textwidth]{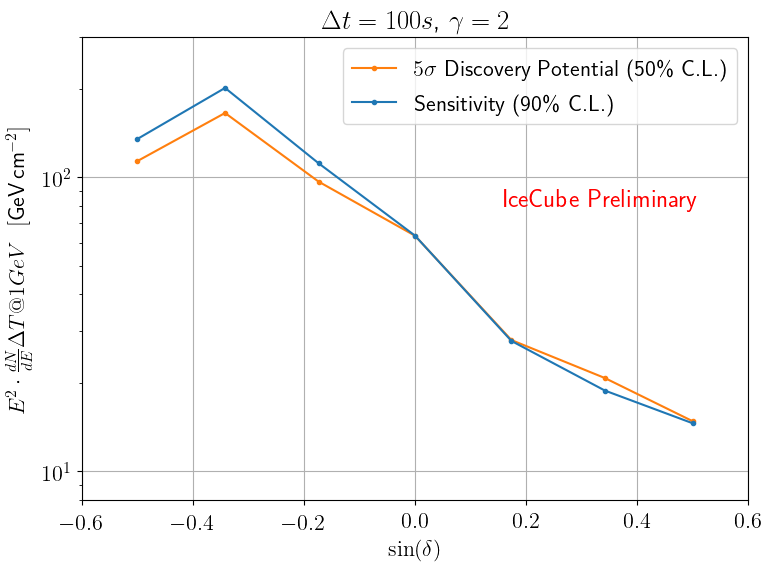}
\caption{Time-integrated sensitivity flux and $5\sigma$ discovery potential flux vs. the sine of declinations for $\Delta T = 100$ seconds. For declinations near poles, results are not reliable due to large angular uncertainties, and thus, half of the whole sky is evaluated. The $5\sigma$ background $TS$ is estimated using the Wilk's theorem.}
\label{fig:sens_dis_dec_dt100}
\end{figure}

As shown in Figure \ref{fig:TS_dist}, the number of events needed to make 50\% of the signal $TS$ distribution larger than the chosen threshold p-value is defined as the `discovery potential'. Similarly, the sensitivity is defined to be the required number of events for 90\% of the signal $TS$ distribution to exceed the median of the background distribution. In this analysis, different injected time windows at different declinations are tested. The sensitivity and $5\sigma$ discovery potential for different flare widths are shown in Figure \ref{fig:sens_dis_dt}. The sensitivities and discovery potentials to 100 s flares at different declinations are shown in Figure \ref{fig:sens_dis_dec_dt100}.

In figure \ref{fig:sens_dis_dt}, the sensitivity and discovery potential become similar at short time widths as signal $TS$ distribution shown in figure \ref{fig:TS_dist} becomes wide with a long tail for a shorter time width. This makes the median line towards a larger $TS$ and results in a low discovery potential. As IceCube has uniform sensitivities in the right ascension, only the declination matters for potential sources. Figure \ref{fig:sens_dis_dec_dt100} shows that the analysis is more sensitive to the transient sources that have flare width of 100 s in the Northern Hemisphere. The declination dependence is related to declination dependence of the effective area as shown in \cite{icrc2021mlarson}. As the effective area is larger for events coming from the Northern Hemisphere, the required fluence emitted from an astrophysical source is smaller in order for the source to be detected. 

\section{Future Searches for GRBs}
Work presented here serves as preparation to search for 10-100 GeV neutrino emission from GRBs. Unlike what we have presented here, GRB searches are triggered analysis, in which $t_0$ and $\sigma_w$ are not fitted. We are studying GRBs deteced between April 2012 - June 2019 that are away from the poles due to the large event angular uncertainties. For the on-going searches, both the subphotospheric model and the choked-GRB model will be tested.

As the directional and triggered temporal information becomes known, only the number of events, spectral index and flare time width will be fitted in the maximum likelihood method. A binomial test described in \cite{aartsen2019search} can then be conducted to evaluate the significance of p-values for the best fit from all potential GRBs. The binomial probability

\begin{equation}
P_{binom}(k|p_k, N) = \binom{N}{k}p_k^k(1-p_k)^{N-k}
\end{equation}
is the probability for no more than $k$ out of $N$ total GRBs to have $p > p_k$, where $p_k$ is the p-value of the $k$th GRB among a list of GRBs ordered by p-values. The best binomial p-value will be obtained at a $k_f$, then the top $k_f$ significant p-values associated the most contributing GRBs will be reported in the future.

\section{Conclusion}

We present an all-sky untriggered search for astrophysical transients with low energy neutrinos observed by IceCube-DeepCore. Sensitivies and discovery potentials for flares with different livetime at different declinations using this method are shown. Our results show that it is possible to search for astrophysical sources like choked-GRBs with this newly developed technique.

\bibliographystyle{ICRC}
\bibliography{skeleton}




\clearpage
\section*{Full Author List: IceCube Collaboration}




\scriptsize
\noindent
R. Abbasi$^{17}$,
M. Ackermann$^{59}$,
J. Adams$^{18}$,
J. A. Aguilar$^{12}$,
M. Ahlers$^{22}$,
M. Ahrens$^{50}$,
C. Alispach$^{28}$,
A. A. Alves Jr.$^{31}$,
N. M. Amin$^{42}$,
R. An$^{14}$
% ...
\\

\noindent
% ...
$^{12}$ Universit{\'e} Libre de Bruxelles, Science Faculty CP230, B-1050 Brussels, Belgium \\
% ...
$^{14}$ Department of Physics and Laboratory for Particle Physics and Cosmology, Harvard University, Cambridge, MA 02138, USA \\
% ...
$^{17}$ Department of Physics, Loyola University Chicago, Chicago, IL 60660, USA \\
$^{18}$ Dept. of Physics and Astronomy, University of Canterbury, Private Bag 4800, Christchurch, New Zealand \\
% ...
$^{22}$ Niels Bohr Institute, University of Copenhagen, DK-2100 Copenhagen, Denmark \\
% ...
$^{28}$ D{\'e}partement de physique nucl{\'e}aire et corpusculaire, Universit{\'e} de Gen{\`e}ve, CH-1211 Gen{\`e}ve, Switzerland \\
% ...
$^{31}$ Karlsruhe Institute of Technology, Institute for Astroparticle Physics, D-76021 Karlsruhe, Germany  \\
% ...
$^{42}$ Bartol Research Institute and Dept. of Physics and Astronomy, University of Delaware, Newark, DE 19716, USA \\
% ...
$^{50}$ Oskar Klein Centre and Dept. of Physics, Stockholm University, SE-10691 Stockholm, Sweden \\
% ...
$^{59}$ DESY, D-15738 Zeuthen, Germany \\
% ...

\subsection*{Acknowledgements}

\noindent
USA {\textendash} U.S. National Science Foundation-Office of Polar Programs,
U.S. National Science Foundation-Physics Division,
U.S. National Science Foundation-EPSCoR,
Wisconsin Alumni Research Foundation,
Center for High Throughput Computing (CHTC) at the University of Wisconsin{\textendash}Madison,
Open Science Grid (OSG),
Extreme Science and Engineering Discovery Environment (XSEDE),
Frontera computing project at the Texas Advanced Computing Center,
U.S. Department of Energy-National Energy Research Scientific Computing Center,
Particle astrophysics research computing center at the University of Maryland,
Institute for Cyber-Enabled Research at Michigan State University,
and Astroparticle physics computational facility at Marquette University;
Belgium {\textendash} Funds for Scientific Research (FRS-FNRS and FWO),
FWO Odysseus and Big Science programmes,
and Belgian Federal Science Policy Office (Belspo);
Germany {\textendash} Bundesministerium f{\"u}r Bildung und Forschung (BMBF),
Deutsche Forschungsgemeinschaft (DFG),
Helmholtz Alliance for Astroparticle Physics (HAP),
Initiative and Networking Fund of the Helmholtz Association,
Deutsches Elektronen Synchrotron (DESY),
and High Performance Computing cluster of the RWTH Aachen;
Sweden {\textendash} Swedish Research Council,
Swedish Polar Research Secretariat,
Swedish National Infrastructure for Computing (SNIC),
and Knut and Alice Wallenberg Foundation;
Australia {\textendash} Australian Research Council;
Canada {\textendash} Natural Sciences and Engineering Research Council of Canada,
Calcul Qu{\'e}bec, Compute Ontario, Canada Foundation for Innovation, WestGrid, and Compute Canada;
Denmark {\textendash} Villum Fonden and Carlsberg Foundation;
New Zealand {\textendash} Marsden Fund;
Japan {\textendash} Japan Society for Promotion of Science (JSPS)
and Institute for Global Prominent Research (IGPR) of Chiba University;
Korea {\textendash} National Research Foundation of Korea (NRF);
Switzerland {\textendash} Swiss National Science Foundation (SNSF);
United Kingdom {\textendash} Department of Physics, University of Oxford.

\end{document}